# An Analysis of XML Compression Efficiency

Christopher J. Augeri[1]  Barry E. Mullins[1]
Dursun A. Bulutoglu[2]  Rusty O. Baldwin[1]

[1]Department of Electrical and Computer Engineering
[2]Department of Mathematics and Statistics
Air Force Institute of Technology (AFIT)
Wright Patterson Air Force Base, Dayton, OH

{chris.augeri, barry.mullins}@afit.edu

{dursun.bulutoglu, rusty.baldwin}@afit.edu

Leemon C. Baird III

Department of Computer Science
United States Air Force Academy (USAFA)
USAFA, Colorado Springs, CO

leemon.baird@usafa.edu

## ABSTRACT
XML simplifies data exchange among heterogeneous computers, but it is notoriously verbose and has spawned the development of many XML-specific compressors and binary formats. We present an XML test corpus and a combined efficiency metric integrating compression ratio and execution speed. We use this corpus and linear regression to assess 14 general-purpose and XML-specific compressors relative to the proposed metric. We also identify key factors when selecting a compressor. Our results show XMill or WBXML may be useful in some instances, but a general-purpose compressor is often the best choice.

## Categories and Subject Descriptors
E.4 [**Data**]: Coding and Information Theory—*Data Compaction and Compression*; H.3.4 [**Systems and Software**]: *performance evaluation (efficiency and effectiveness)*

## 1. INTRODUCTION
Statistical methods are often used for analyzing experimental data; however, computer science experiments often only provide a comparison of means. We describe how we used more robust statistical methods, i.e., linear regression, to analyze the performance of 14 compressors against a corpus of XML files we assembled with respect to an efficiency metric proposed herein.

Our end application is minimizing transmission time of an XML file between wireless devices, e.g., nodes in a distributed sensor network (DSN), for example, an unmanned aerial vehicle (UAV) swarm. Thus, we focus on compressed file sizes and execution times, foregoing the assessment of decompression time or whether a particular compressor supports XML queries.

This paper is authored by employees of the U.S. Government and is in the public domain. This research is supported in part by the Air Force Communications Agency. The views expressed in this paper are those of the authors and do not reflect the official policy or position of the U.S. Air Force, Department of Defense, or the U.S. Government.
*ExpCS*, 13–14 June 2007, San Diego, CA
978-1-59593-751-3/07/06

We expand previous XML compression studies [9, 26, 34, 47] by proposing the XML file corpus and a combined efficiency metric. The corpus was assembled using guidelines given by developers of the Canterbury corpus [3], files often used to assess compressor performance. The efficiency metric combines execution speed and compression ratio, enabling simultaneous assessment of these metrics, versus prioritizing one metric over the other. We analyze collected metrics using linear regression models (ANOVA) versus a simple comparison of means, e.g., $X$ is 20% better than $Y$.

## 2. XML OVERVIEW
XML has gained much acceptance since first proposed in 1998 by the World-Wide Web Consortium (W3C). The XML format uses schemas to standardize data exchange amongst various computing systems. However, XML is notoriously verbose and consumes significant storage space in these systems. To address these issues, the W3C formed the Efficient XML Interchange Working Group (EXIWG) to specify an XML binary format [15]. Although a binary format foregoes interoperability, applications such as wireless devices use them due to system limitations.

### 2.1 XML Format
The example file shown in Figure 1 highlights the salient features of XML [17], e.g., XML is case-sensitive. A *declaration* (line 1) specifies properties such as a text encoding. An *attribute* (line 3) is similar to a variable, e.g., 'author="B. A. Writer"'. A *comment* begins with "<!--" (lines 4, 8). An *element* consists of the elements, comments, or attributes between a tag pair, e.g., "<Chapter>" and "</Chapter>" (lines 5–7). An example of an XML *path* is "/Book/Chapter". A well-formed XML file contains a single *root* element, e.g., "Book" (lines 2–12).

```
 1  <?xml version="1.0" encoding="UTF-8"?>
 2  <Book><Title>Bestseller</Title>
 3  <Info author="B. A. Writer"></Info>
 4     <!-- Write early, write often -->
 5     <Chapter><Title>Plot begins</Title>
 6       <Par>...dark and stormy...</Par>
 7     </Chapter>
 8     <!-- ... -->
 9     <Chapter><Title>Plot ends</Title>
10       <Par>...antagonist destroyed!</Par>
11     </Chapter>
12  </Book>
```

**Figure 1. XML sample file**

## 2.2 XML Schemas

An XML schema specifies the structure and element types that may appear in an XML file and can be specified in three ways: implicitly by a raw XML file or explicitly via either a document type definition (DTD) or an XML Schema Definition (XSD) file. A schema for Figure 1 can be implicitly obtained via the path tree defined by its tags. However, implicit schemas do not enable the data validation possible by explicitly declaring a DTD or XSD.

The use of external schemas may result in smaller XML files and also enables data validation. The least robust schema approach is a DTD, whereas an XSD schema is itself an XML file. Certain compressors *require* a schema file—we generated DTDs to enable us to test these compressors (cf. Section 5.1). Further discussion of XML is beyond the scope of this paper, however, we note two parser models, SAX and DOM, are often used to read and write XML files. Other standards, e.g., XPath, XQuery, and XSL, provide robust mechanisms to access and query XML data.

## 3. INFORMATION AND COMPRESSION

We measure entropy as classically defined by Shannon [37] and replicated here, using slightly modified notation, where,

$$H_n \stackrel{def}{=} -\frac{1}{n} \cdot \sum_{\forall s, s \in S_n} \left\{ \Pr(s) \cdot \log_2 \left[ \Pr(s) \right] \right\}, \quad (1)$$

$S_n$ denotes all words having $n$ symbols, $\Pr(s)$ the probability that a word, $s \in S_n$, occurs, and $H_n$ the entropy in bits per symbol.

For example, if $S_1$ contains {A, B}, all 2-word combinations of $S_1$ yields an $S_2$ of {AA, AB, BA, BB}, $S_3$ is {AAA, AAB, …, BBB}, and so forth. The true entropy, $H_\infty$, is obtained as $n$ grows without bound. If all words are equally probable and independent, the data is not theoretically compressible; most real-world data is not random, thus, some data compression is usually obtainable.

An optimal lossless compressor cannot encode a source message containing $m$ total symbols at less than $m \cdot H_\infty$ bits on average. Compressors cannot achieve optimal compression since perfectly modeling a source requires collecting an infinite amount of data. In practice, compressors are limited by the rapidly increasing time and space requirements needed to track increasing word lengths.

The first-order Shannon entropy, $H_1$, corresponds with a 0-order Markov process; herein, we use both as appropriate. Simply put, a 1-order compressor can track word lengths of two symbols and an 11-order compressor can track words lengths of 12 symbols. This can often be confusing, as some compressors use 1-byte character symbols and others may use multi-byte words. We thus can view a 7-order 1-bit compressor as a 0-order 1-byte compressor.

The compression process model used herein is shown in Figure 2. A *lossless* transform is a pre-processing method that attempts to reduce $H_\infty$ by ordering data in a canonical form. A *lossy* transform reduces $m$, and in doing so, may reduce $H_\infty$. Lossy and lossless transform(s) are optional, denoted by a shaded box. The compression step attempts to store the data in a format using less bits than the native format, i.e., it attempts to store the data at $H_\infty$.

A lossless transform, e.g., Burrows-Wheeler (cf. Section 4.1.2.3) sends *at least* the original number of bits to the compressor. A lossy transform, such as is used in the MP3 and JPEG file formats, discards "extra" bits. Lossy transforms are not typically applied to textual or numerical data, i.e., although humans can often compensate for missing pixels (frequencies), it is often difficult to guess absent characters (numbers). Following any pre-processing transforms, a lossless compressor is applied; although most files are smaller after this step, some files *must* be larger. This effect may be observed when compressing small files, previously compressed files, or encrypted files. All lossless compression steps are reversed by decompression.

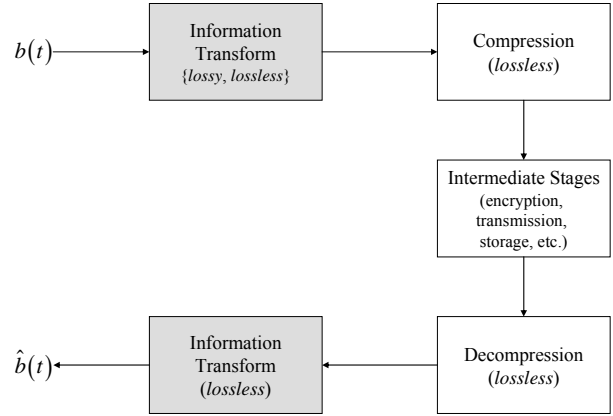

**Figure 2. Compression and decompression pipeline stages**

## 4. COMPRESSORS

General-purpose compressors can be classified within two classes, arithmetic or dictionary. Arithmetic compressors typically have large memory and execution time requirements, but are useful to estimate entropy and as control algorithms. The dictionary, or zip, compressors enjoy widespread use and most have open formats. Some proprietary formats are 7-zip, CAB, RK, and StuffIt [13] and may yield smaller files than the zip compressors. We note the *primary* criterion for inclusion of an XML compressor within this study is whether a *publicly* accessible implementation is available. Our primary objective was to test any available XML compressor, as evidenced by our use of Internet archives and a Linux emulator to enable us to test certain compressors (cf. Appendices B and C).

### 4.1 General-Purpose

#### 4.1.1 Arithmetic Compressors

An arithmetic compressor estimates the probability of a symbol using a specific buffer and symbol length. Although floating-point numbers are often used to explain arithmetic compression, most implementations use integers and register shifts for efficiency.

An arithmetic compressor uses a static or dynamic model. A static model can either be one based on historical data or generated *a priori* before the data is actually encoded. If a compressor uses a dynamic model, the model statistics are updated as a file is being compressed; this approach is often used when multiple entropy models are being tracked, such as in the PPM compressors. Given the computational power required, however, dynamic models have only recently been used in practice.

Since the CACM3 compressor [33] closely approximates $H_1$, we used it to validate our computation of $H_1$ and to estimate the worst expected compression ratio. The other arithmetic compressors, PAQ [28] and PPM [7], provided an estimate, $E[H_\infty]$, on the entropy bound, $H_\infty$, achievable for each file, i.e., we used $E[H_\infty]$ to estimate the maximum expected compression of each test file.

*4.1.1.1 CACM3*
CACM3, or the Communications of the ACM compressor [33], is well-known and is often used as a reference model for 0-order arithmetic compression. This compressor uses three stages: model updating, statistics gathering, and encoding. For example, the model specifies if a 0-order character-based model or a multiple-context 0-order word-based model is being used. Having multiple contexts enables the compressor to maintain statistics for different data types, e.g., binary versus textual data.

As expected, CACM3 closely approximates the 0-order entropy, e.g., given several copies of the sentence, "*The quick brown fox jumps over a lazy dog.*" CACM3 encodes it using 4.494 bits/byte, within 1.22% of the true 0-order entropy, $H_1$, 4.440 bits/byte. We used CACM3 as a control algorithm to validate the computed $H_1$. CACM3 was within 10% of $H_1$ of files larger than 1 KB (42 files) and within 1% of $H_1$ of files greater than 64 KB (22 files).

*4.1.1.2 PPM*
In the limit, Prediction by Partial Matching (PPM) compressors are considered theoretically optimal [7, 32]. The goal of PPM is to match the maximum possible symbol length; if a new symbol is encountered, a special "escape" symbol is inserted, and a symbol encoding is generated for the new symbol. Newer PPM variants, such as PPMZ2 [4], use multiple contexts and encoders, along with improved escape symbol encoding, to improve compression performance.

*4.1.1.3 PAQ*
PAQ is the third arithmetic encoder and is best identified as a PPM hybrid [28]. Although PAQ is a PPM variant, it may use other techniques, e.g., run-length encoding (RLE), if it determines these encodings yield more compression. PAQ is an open-source compressor and several people contribute to its development. The version used in this study is PAQsDaCC 4.1.

*4.1.2 Dictionary Compressors*
All of the commonly-used zip formats are at least partially based on dictionary compression algorithms and have their roots in the Lempel-Ziv (LZ77) algorithm [51]. At least one zip compressor is typically used as a reference compressor in compression studies or when presenting a new compression algorithm. We assessed three common variants: WinZip® [45], GZIP [19], and BZIP [36].

*4.1.2.1 WinZip®*
The WinZip® compressor [45] uses the *deflate* format (RFC 1951) by default. In version 10.0, BZIP2 [36] and PPMd [38] support were also added. As its name suggests, PPMd is a PPM-based compressor and often executes faster than other PPM. We tested WinZip® in Deflate, Enhanced Deflate, and PPMd modes; the PPMd variant was assessed using an evaluation version.

*4.1.2.2 GZIP*
GZIP [19] is a variant of the LZ77 algorithm [51], where symbol combination match lengths are stored in a Huffman tree; another Huffman tree stores match distances, where a match distance is the distance between consecutive symbol combinations. The user has the option of setting the maximum match length and match distances to compare. The maximum match length in GZIP is 258 bytes and the maximum match distance is 32 KB.

*4.1.2.3 BZIP2*
The BZIP2 compressor [36] uses the lossless Burrows-Wheeler transform (BWT) [5]. The BWT groups identical symbols by sorting all possible rotations. For example, the text ".BANANA_" becomes "BNN.AA_A" [6], where '_' denotes the end-of-file (EOF) symbol. The BWT exposes the true entropy of the original file at lower Markov orders. Given the BWT was designed for text data, BZIP2 can be expected to perform well on XML data. The BWT is also effective on data other than plaintext characters.

## 4.2 Binary XML Formats
Binary formats encode XML documents as binary data. The intent is to decrease the file size and reduce the required processing at remote nodes. However, a binary format runs counter to the key benefit of using XML—interoperability. If XML binary formats are to succeed, an open standard must be established.

The primary impetus for binary XML is the limited capabilities of wireless devices, e.g., cell phones and sensor networks. Further pressure to use a binary format comes from the growth of large repositories, e.g., databases that store data using an XML format. Technically, both compressed and binary formats are "binary" formats, versus plaintext, but binary formats may support random access and queries, whereas compression formats often do not.

*4.2.1 W3C Initiatives*
The XML-binary Optimized Packaging (XOP) supports the inclusion of binary data, e.g., image and sound files, in XML. The XOP enabling mechanism is *base64*, the encoding used to send an attachment via e-mail, and increases the data's size by a 4:3 ratio. Conversely, a "pure" binary format converts XML (including "packaged" binary data!) to binary to reduce its storage footprint.

*4.2.2 WBXML*
The Wireless Binary XML (WBXML) format achieves two goals: compression of XML and stream-level processing, and as its name implies, was developed to support wireless devices. By reducing file size, WBXML addresses a key power management issue in mobile devices, while providing many of XML's benefits and incurring minimal overhead. Additionally, a schema can be pre-loaded on the wireless device, further reducing the size of the transmitted XML file. We used an existing converter [44] that yields a WBXML-formatted file in this study.

*4.2.3 XBIS*
XBIS is a binary XML encoding format and stream encoder that receives SAX events during decompression and also retains the XML schema format of the native data [46]. XBIS influenced the development of ASN.1; the implementation we used [46] also was the source of two test corpus files.

*4.2.4 ASN.1 / Fast Infoset*
Abstract Syntax Notation One (ASN.1) is an International Telecommunications Union (ITU) standard (X.891) predating XML and WBXML. There are ways of translating between XML and ASN.1, e.g., IBM's ASN.1/XML translator. ASN.1 also has support for binary versus plain-text encodings [1]. The Fast Infoset (FIS) specification defines how to translate from XML to binary using ASN.1 and is being modified to provide data security features. Sun Microsystems is extending its Java programming language to support FIS and is the variant we assessed [18].

## 4.3 Schema-Aware XML Compressors

### 4.3.1 XGRIND
XGrind is an early XML-specific compressor [42]. The resulting files may be queried, thus, it may be viewed as a binary formatter. XGrind compresses the XML and schema files to separate "*.xgr" and "*.met" files, respectively. Other compressors were tested in Microsoft Windows®, however, we had to use a Linux emulator, CoLinux (cf. Appendix B), to test the available XGrind variant.

### 4.3.2 XMLPPM
XMLPPM [7] is based on a PPM algorithm. XMLPPM provides a streaming access model (ESAX) and multiplexed hierarchical PPM (MHM) models. The combined goal is to increase parsing by using ESAX and compression performance by using MHM. MHM uses different, i.e., multiple, PPM contexts depending on whether tags, attributes, or elements are being currently encoded. A DTD-based variant of XMLPPM has recently been developed. The newest versions use PPMd as the internal PPM compressor.

### 4.3.3 XMILL
The XMill compressor applies a pre-processing transform and then uses GZIP compression [27]. The pre-processing transform separates the XML tree structure according to element names into context-specific containers; the XML tree structure and containers are then passed to GZIP. By default, each unique element name specifies a unique container; users can also specify containers. Containers are individually compressed by GZIP on the premise that data in commonly named elements may have similar entropy. This extended discussion of XMill is given based on the large set of user options XMill possesses relative to the other compressors.

The default XMill usage is shown in row 1 of Table 1. The "//#" specifies each unique element is mapped to one container. All XML paths sharing the same trailing element name are placed in the same container. Thus, the three paths "/db/car/color", "/db/paint/color", and "/db/fruit/color" would have their data values placed in one container. Again, this is under the premise "color" elements may compress better if together.

Using the command in row 2 maps "/db/car/color" to its own container. All other elements are grouped into last-element containers, e.g., the "/db/fruit/color" and "/db/color" paths would both be placed in the "color" element container. The "-p //#" option is only listed for illustration; it is implicitly added by XMill to provide a default container for all elements.

To list all XML elements in a file, use the command on row 3. To force XMill to map all elements to the same container, the command in row 4 should be used, where "t1" represents the first element and "tn" is the last element. Quotes ('"') are needed when a pipe ('|') is used or it is treated as a command-line pipe. A simpler approach to mapping all elements to the same container is to use the command given in row 5, as each element does not have to then be explicitly identified. There is also a way to map each unique path to its own container, shown in row 6.

XMill provides a set of compressors to use before compression by GZIP. The default compressor is a non-compressing pass-through compressor. However, a container's compressor can be specified by the user, e.g., row 7 uses a run-length encoder included with XMill. Other proprietary compressors are also provided with XMill or others can be linked by the user via command-line.

**Table 1. XMill user command-line option examples**

| ID | XMill Command-Line Example |
|---|---|
| 1 | `xmill –p //# dat.xml` |
| 2 | `xmill –p //db/car/color –p //# dat.xml` |
| 3 | `xmill –v dat.xml` |
| 4 | `xmill –p "//(t1 | t2 | ... | tn)" dat.xml` |
| 5 | `xmill –p "//(*)" dat.xml` |
| 6 | `xmill –p (#)+ dat.xml` |
| 7 | `xmill –p "//(*)=>rl" dat.xml` |
| 8 | `xmill –p "//(#)=>t" dat.xml` |

Row 8 explicitly describes XMill's default behavior: each unique element is mapped to a container and the default pass-through text compressor is used. There are also additional options available in XMill; this level of customization is a unique feature of XMill.

### 4.3.4 XML-ZIP
XML-ZIP is an early XML-specific compressor that divides the XML tree into a set of sub-tree files [50] and was discussed in at least one earlier study [34]. The splitting depth parameter used in XML-ZIP specifies that branches below the splitting depth are to be compressed using zip compression. The top-level branches are not compressed, but stored in their native form. All sub-tree fragments are listed as a sub-tree file in the zip archive, where the zip archive contains the following sub-files:

1. Encoded XML file of the original tree (text)
2. One or more nested sub-tree files (compressed)
3. Compressed element file mapping (text)

To apply XML-ZIP to an XML file, all comments should first be removed and all tags listed one per line (a reformat in Microsoft FrontPage® obtains the latter), e.g., as shown in Figure 1. If a splitting depth of '2' is specified, all branches at or below "Book" would be individually compressed and the book title and chapters are contained in separate sub-tree file fragments.

This process yields an encoded XML tree; the tree derived from applying a splitting depth of '2' to Figure 1 is shown in Figure 3. This can be thought of as a dictionary look-up, where an element, e.g., "`<xmlzip id="2"/>`", is a single file in the archive. The mapping of elements to files is given separately (item 3 above). In this example, four compressed element files are in the zip archive, along with the encoded XML tree and file mapping. XML-ZIP is akin to XMill in that it applies a lossless pre-processing transform and then uses a general-purpose compressor for compression.

```
1  <?xml version="1.0" encoding="utf-8"?>
2  <Book><Title><xmlzip id="1"/></Title>
3    <Info author="B. A. Writer">
4      <xmlzip id="2"/>
5    </Info>
6    <Chapter><xmlzip id="3"/>Chapter>
7    <Chapter><xmlzip id="4"/>Chapter>
8  </Book>
```
**Figure 3. XML-ZIP encoding tree example**

## 4.4 Additional Compressors (Non-Tested)

We recall our key criterion for testing an XML compressor herein is access to a publicly available implementation; the compressors in this section did not have such an implementation available. Some of these are analyzed in previous studies [9, 34, 47], such as AXECHOP [24], BOX [2], Millau [20, 41], and XCOMP [26]. We attempted to include Efficient XML [16] in this study, but did not receive a response to our requests to do so. However, since the time of this study, the EXIWG selected Efficient XML as the basis of their XML binary format specification [15].

### 4.4.1 XML-Xpress
Intelligent Compression Technologies developed a compressor, XML-Xpress [49] that requires a vendor-provided Schema Model File (SMF). An SMF is a static statistics model tailored to a set of similar XML files; this technique may approach the performance of an arithmetic compressor but is difficult to scale. We used the ham radio files shipped with XML-Xpress in our test corpus.

### 4.4.2 XPRESS
The XPRESS algorithm [31] introduces the concept of reverse arithmetic encoding (RAE). In RAE, the entire XML hierarchy is mapped over the real interval [0.0, 1.0). For instance, "/Book" would be assigned to the range [0.0, 1.0). The sub-element "/Book/Info" may potentially map to the range [0.0, 0.2) and the element "/Book/Info/author" to [0.0, 0.15). XPRESS supports queries of the resulting file without full decompression.

### 4.4.3 MPEG-7 (BiM)
The binary format for MPEG-7 can encode XML and is designed for streaming media [9]. For example, MPEG-7 can take metadata and encode it in binary, e.g., to facilitate closed-captioning. The MPEG-7 format is also referred to as BiM.

### 4.4.4 Proprietary Formats
Oracle, IBM, and Microsoft each offer proprietary binary XML formats in their respective database server products. The Open Geospatial (OpenGIS) Consortium defined a binary XML format, B-XML, based on the Geography Markup Language. CubeWerx has implemented B-XML as CWXML; the XML files packaged with CWXML were included in our test corpus.

## 5. METHODOLOGY

### 5.1 Test Files
Although several general-purpose compression corpora exist, an XML test corpus does not yet exist. The W3C, in conjunction with the National Institute of Standards and Technology (NIST), is developing XML *conformance* tests. However, little work has been done to develop an XML *performance* corpus. Although benchmarks do not completely describe a system's performance, they do provide for consistency in the literature. This is especially true if a benchmark, in this case, the proposed XML test corpus, is crafted to represent a cross-section of source domains and system demands, as shown in Table 2.

The corpus was assembled based on recommendations given by the designers of the oft-cited Canterbury corpus [3]. For example, files were chosen based on topical domain coverage, raw file size, similarity to other corpora, e.g., the Canterbury [3] or Calgary corpora [14], and their public availability, where source locations of each file is provided in Appendix C.

A subset of the factors collected from the files is also given in Table 2. The test files used ranged from less than 1 KB to 4 MB, and one large ~40 MB file. Since parser quality varied among the compressors tested, some preliminary processing of the corpus files was necessary prior to beginning the experiments. The pre-processing performed on each file is as follows:

1. **Validation and Beautification**: *Tidy* is an XML and HTML validation and beautification tool [35]. It was used to remove blank lines and to indent each level for readability. Tidy also identifies certain validation errors; those found were corrected before proceeding.

2. **Schema Extraction**: Although DTD and/or XSD files were often available with the original XML files, some schemas contained errors (as reported by Microsoft FrontPage®). Since some compressors require a schema, we had to define one for every test file. To minimize test errors, we generated a schema for each file versus using any provided schema(s) or schemas of any similar files. We removed references to any existing schemas in a file and generated an explicit schema for it by applying an open-source Java-based implicit schema extractor [23].

3. **Schema Cross-Check**: Microsoft FrontPage® was used to validate the XML file yielded by *tidy* and the extracted implicit schema. We removed all comments, since some compressors did not properly parse them, and added a reference to the extracted schema, or DTD.

4. **XML Factors**: The only XML statistics calculator we located was a PHP-based statistics package [48] and that was slightly modified for our use. This package was used to collect various properties of a file, e.g., its XML tree depth. The PHP parser had some problems with blank lines and comments; however, these issues were resolved in the course of steps 1-3.

5. **Line Count**: This property is not provided by the script in step 4. We gathered it via the command-line by using "`type foo.xml | find /v /c ""`".

6. **Zero-Order Entropy ($H_1$)**: We used another PHP script to calculate 0-order entropy [30]. Since XML only contains text, we assume each symbol is one byte, i.e., one ASCII character. Thus, we used the 1-byte Unicode encoding (UTF-8, similar to ASCII) to save corpus files.

7. **Entropy Estimate ($E[H_\infty]$)**: The estimate of true entropy was based on the best compression achieved for each file after executing the PAQ and PPM compressors at their maximum compression settings.

The factors listed in Table 2 were used to fit our linear regression models. The file and description columns identify the file name and a description of the file's origin. The file domain is a subjective classification that will be explained shortly. The next five columns list the uncompressed file size (in bytes), number of lines, number of unique characters (1-byte symbols), number of unique tags, and the XML tree depth (as defined by the XML tags contained in each file). The value for $H_1$ is given by step 6 above and validated by CACM3; the value for $E[H_\infty]$ is the result of the best compression obtained for that file (typically PAQ) after being compressed by every compressor tested.

Table 2. Corpus test file properties (cf. Appendix C)

| File Name | Description | Domain | Bytes | Lines | Chars | Tags | Depth | $H_1$ | $E[H_\infty]$ |
|---|---|---|---|---|---|---|---|---|---|
| AB_FR_META | Weather Data (2004): France, Norway, Turkey | SI | 40042243 | 841765 | 85 | 88 | 10 | 3.987 | 0.008 |
| AB_NO_META | | SI | 1171129 | 24652 | 83 | 85 | 10 | 3.989 | 0.011 |
| AB_TR_META | | SI | 335 | 9 | 58 | 5 | 3 | 5.248 | 0.325 |
| BB_1998STATS | Baseball Stats (1998) | DB | 904261 | 25965 | 76 | 43 | 6 | 4.373 | 0.020 |
| CB_CONTENT | OpenDocument Sample File | MU | 814397 | 17714 | 94 | 35 | 11 | 4.890 | 0.046 |
| CB_WMS_CAPS | GIS Map Server Data | DB | 1004047 | 18557 | 87 | 35 | 10 | 4.849 | 0.035 |
| LW_H2385_RH | US House Of Representatives: Bill Resolutions | LW | 5337 | 115 | 79 | 35 | 5 | 5.080 | 0.247 |
| LW_H3738_IH | | LW | 5167 | 118 | 78 | 25 | 8 | 5.046 | 0.276 |
| LW_H3779_IH | | LW | 6336 | 132 | 75 | 27 | 9 | 4.811 | 0.249 |
| LW_ROLL014 | US House Of Representatives: Roll Call Votes | LW | 100499 | 2651 | 82 | 32 | 5 | 4.751 | 0.040 |
| LW_ROLL020 | | LW | 100568 | 2652 | 82 | 32 | 5 | 4.751 | 0.040 |
| LW_ROLL031 | | LW | 100368 | 2655 | 82 | 32 | 5 | 4.756 | 0.040 |
| NT_BOOLEAN | NIST XML Data Type Conformance Tests | NT | 7563 | 142 | 85 | 6 | 6 | 5.127 | 0.157 |
| NT_NORMSTRING | | NT | 62800 | 1193 | 85 | 6 | 6 | 4.822 | 0.028 |
| NT_POSLONG | | NT | 81669 | 1584 | 85 | 6 | 6 | 4.850 | 0.026 |
| OD_ALLEN | Oracle Database Sample Transactions | DB | 2445 | 63 | 75 | 17 | 4 | 5.149 | 0.065 |
| OD_FORD | | DB | 3958 | 99 | 73 | 17 | 4 | 5.072 | 0.039 |
| OD_MILLER | | DB | 4430 | 111 | 76 | 17 | 4 | 5.061 | 0.037 |
| PD_CONNOW | DoD Per Diem Data (2003) | LI | 481983 | 13689 | 77 | 12 | 3 | 5.126 | 0.019 |
| PD_CONUSMIL | | LI | 191796 | 5406 | 77 | 12 | 3 | 5.136 | 0.025 |
| PD_CONUSNM | | LI | 290380 | 8287 | 77 | 12 | 3 | 5.117 | 0.020 |
| PY_AS_YOU | Shakespeare: As You Like It, Comedy Of Errors, Hamlet | MU | 244498 | 6360 | 76 | 18 | 6 | 4.655 | 0.125 |
| PY_COM_ERR | | MU | 170641 | 4160 | 76 | 16 | 6 | 4.728 | 0.125 |
| PY_HAMLET | | MU | 352466 | 8836 | 77 | 16 | 6 | 4.703 | 0.128 |
| RS_AP | RSS "Top Story" News Feeds: AP, CNET, Reuters | RS | 6203 | 120 | 80 | 7 | 4 | 5.057 | 0.247 |
| RS_CNET_SMALL | | RS | 7328 | 177 | 76 | 18 | 4 | 5.112 | 0.183 |
| RS_CNET_LARGE | | DB | 251900 | 5551 | 85 | 13 | 4 | 5.099 | 0.090 |
| RS_REUTERS | | RS | 7868 | 149 | 80 | 12 | 4 | 5.329 | 0.179 |
| WX_29 | NOAA Weather Forecasts (3 locations) | SI | 41460 | 1104 | 73 | 40 | 7 | 4.603 | 0.032 |
| WX_38 | | SI | 26897 | 744 | 73 | 39 | 6 | 4.672 | 0.046 |
| WX_39 | | SI | 26887 | 744 | 73 | 39 | 6 | 4.670 | 0.046 |
| XB_FACTBOOK | CIA World Factbook | BK | 5047775 | 106938 | 87 | 199 | 5 | 4.878 | 0.081 |
| XB_PERIODIC | Periodic Table of Elements | SI | 107147 | 2428 | 78 | 20 | 3 | 5.307 | 0.035 |
| XG_STUDENT | Student Degree Listing | DB | 30411 | 1000 | 74 | 6 | 3 | 5.216 | 0.067 |
| XM_DBLP | Bibliographic Database | DB | 107864 | 2939 | 83 | 18 | 4 | 5.063 | 0.090 |
| XM_SHAKE | Shakespeare: Antony & Cleopatra | MU | 318487 | 8228 | 76 | 17 | 6 | 4.723 | 0.120 |
| XM_SPROT | DNA Sequences | SI | 11946 | 350 | 78 | 28 | 5 | 5.257 | 0.164 |
| XM_TPC | Database Benchmarks | DB | 349213 | 12966 | 77 | 45 | 4 | 4.963 | 0.101 |
| XM_TREEBANK | Wall Street Journal Linguistics | TR | 9129 | 370 | 68 | 28 | 15 | 2.960 | 0.094 |
| XM_WEBLOG | Apache Web Server Log | LI | 2179 | 58 | 75 | 10 | 3 | 5.221 | 0.252 |
| XX_F21000 | FCC Ham Radio Listings | DB | 64055 | 2407 | 70 | 22 | 4 | 4.916 | 0.044 |
| XX_F26000 | | DB | 6573 | 245 | 70 | 22 | 4 | 4.945 | 0.108 |
| XX_F29000 | | DB | 824 | 28 | 66 | 21 | 4 | 5.170 | 0.436 |
| XZ_UNSPSC | UN Product Catalog Code Tree | LI | 1128895 | 31086 | 82 | 6 | 6 | 4.517 | 0.040 |

The test files were grouped within a set of domains, based on our subjective judgment, as listed in Table 3. We use these groupings (cf. third column of Table 2) to determine if the source domain is a significant factor in the linear regression models we develop.

Table 3. Corpus test file domains

| Source Domain | Short Name |
|---|---|
| Books | BK |
| Databases | DB |
| Directory Listings | LI |
| Legal Documents | LW |
| Office Documents | MU |
| Source code (NIST) | NT |
| RSS Feed (News) | RS |
| Scientific Data (Numbers) | SI |
| Deep Nesting (Trees) | TR |

## 5.2 Compressor Configuration and Execution

The compressors are enumerated in Table 4, along with their short names and categorizations based on each compressor's design and description in the literature. The 3-letter short name, e.g., BZ2, was used as part of the file name to archive the files produced by each compressor and to identify axes of relevant figures herein. The compressor class and application columns are used as factors for linear regression model fitting, similar to the file domains.

The compressor class column entries, [ZIP, MAT, XBN, XSC] correspond with the [zip, arithmetic, XML binary, XML schema-aware]-based compressors, respectively. The application column differentiates general-purpose (GLO) and domain-specific (XML) compressors. The subjectivity used to obtain Table 4 is less than that used to designate the test file domains in Table 3, since the categorization in Table 4 is based on a compressor's algorithm.

**Table 4. Compressor classes**

| Compressor | Short Name | Class | Application |
|---|---|---|---|
| BZIP2 | BZ2 | ZIP | GLO |
| CACM3 | CAC | CTL | GLO |
| FIS | FIS | XBN | XML |
| GZIP | GZP | ZIP | GLO |
| PAQ | PAQ | MAT | GLO |
| PPMD | PPD | MAT | GLO |
| PPMZ2 | PPM | MAT | GLO |
| WBXML | WBX | XBN | XML |
| WINZIP | WZP | ZIP | GLO |
| XBIS | XBS | XBN | XML |
| XGRIND | XGR | XSC | XML |
| XMILL1 | XM1 | XSC | XML |
| XMILL2 | XM2 | XSC | XML |
| XMLPPM | XPM | XSC | XML |
| XMLZIP | XZP | XSC | XML |

Each compressor was tested with its default settings; in addition, six compressors, CACM3, GZIP, PAQ, XGrind, XMill, WinZip®, were also tested at their maximum compression settings. The compression execution commands are given in Table 5, where the maximum compression settings, shown in gray, are not used for default compression. Conversely, the "-en" and "H V" options are not used for maximum compression in WinZip® and XGrind.

Any output from the compressors was redirected to archive files. The redirection is omitted, but is of the form "1>>com.1.txt 2>>com.2.txt", where "com" is replaced by the appropriate 3-letter code for a compressor (cf. Table 4). Output redirection, e.g., "1 > dat.com", is explicitly given for CACM3 and GZIP compressors, since the compressed file is provided via standard out (*stdout*) by these compressors.

**Table 5. Compressor execution commands**

| Compressor | Usage |
|---|---|
| BZIP2 | BZIP2 -k -f -v foo.xml |
| CACM3 | arith -e -t word -m 255 -c 20 foo.xml 1 > foo.cac |
| FIS | java -cp FastInfoset.jar com.sun.xml.fastinfoset.tools.XML_SAX_FI foo.xml foo.fis |
| GZIP | GZIP -9 -c -f -v foo.xml 1>foo.gzp |
| PAQ | pasqda -7 foo.paq foo.xml |
| PPMD | wzzip –ep foo.ppd foo.xml |
| PPMZ2 | ppmz2 -e foo.xml foo.ppm |
| WBXML | xml2wbxml -k -o foo.wbx foo.tdy |
| WINZIP | wzzip –en –ee foo.wzp foo.xml |
| XBIS | java -Dorg.xml.sax.driver= com.bluecast.xml.Piccolo -cp Piccolo.jar;saxxbis.jar;.; test.RunTest XBIS foo.xml |
| XGRIND | ./compress foo.tdy H V A N |
| XMILL1 | xmill -m 470 -9 -f -v -w foo.xml |
| XMILL2 | xmill -f -v -w -p "//(*)" foo.xml |
| XMLPPM | xmlppm foo.xml foo.xpm |
| XMLZIP | java -cp xml4j.jar;. XMLZip foo.xml 2 |

## 5.3 Metrics

The metrics collected from the compressors were the compression execution time and the compressed file size. Based on analysis of the residuals and data distributions, transforms were needed to fit a linear regression model. For all transforms, the logarithm is used to linearly distribute the metric; its effect is most observable when comparing execution times. A small value, '+$b$', is used to shift each transform, where $b$ is the logarithm base used in a transform; this biases a transform to positive values that are away from and greater than zero. The biasing minimizes the chances of using a numerically ill-conditioned model matrix for linear regression. In the compression ratio transform,

$$y_{\text{comp\_ratio}} = \log_2\left[8 \cdot s_{\text{native}}/s_{\text{comp}}\right] + 2, \quad (2)$$

we first convert file sizes to bits, where the native file size, $s_{\text{native}}$, is in the numerator to minimize the possibility of encountering a negative logarithm. In the compressor execution speed transform,

$$y_{\text{comp\_speed}} = \log_{10}\left[8 \cdot s_{\text{native}}/t_{\text{exec}}\right] + 10, \quad (3)$$

we divide file size by execution time, where $s_{\text{native}}$ is in bytes and execution time, $t_{\text{exec}}$, is continuously distributed over one 24-hour day, (0.0, 1.0), e.g., if the execution of a run takes 12 hours, then $t_{\text{exec}} = 0.5$. A greater value reflects a higher compression ratio or execution speed in (2) and (3), respectively.

A key goal was to devise and assess a combined efficiency metric. We used a slightly modified version of the only known combined efficiency metric [40] as a control metric,

$$y_{\text{eff\_old}} = \log_{10}\left[2^{(s_{\text{comp}}/\min(s_{\text{comp}})-1)} \cdot t_{\text{exec}}\right] + 10, \quad (4)$$

where $s_{\text{comp}}$ and $t_{\text{exec}}$ are the same values used in (2) and (3). We denote the minimum size across all compressors for a given test file by $\min(s_{\text{comp}})$. Unfortunately, although (4) combines space and time, it yields poor residual plots and normal plots, along with a low $R^2$ value when fitted to linear models. Unlike other metrics herein, a smaller value for (4) is better. Based on our observations with respect to (4), we propose a new combined efficiency metric,

$$y_{\text{eff\_prop}} = \log_{10}\left[\frac{\left(\frac{s_{\text{native}}^2}{\left(\min(s_{\text{comp}}) \cdot s_{\text{comp}}\right)}\right)}{t_{\text{exec}}}\right] + 10, \quad (5)$$

where the exponential function is now eliminated and $t_{\text{exec}}$ is in the denominator. The numerator is a product of the best compression ratio obtained for each test file, $s_{\text{native}}/\min(s_{\text{comp}})$, and that of a compressor, $s_{\text{native}}/s_{\text{comp}}$, in order to standardize the results. Thus, this metric can only be computed after all test runs are executed, since $\min(s_{\text{comp}})$ is the minimum of all compressors. A higher value for $y_{\text{eff\_prop}}$ reflects a better efficiency with respect to compression ratio and execution speed for a given compressor.

In (5), we first included all aspects of compression performance, i.e., compression ratio and execution time. In addition, we wanted to improve on defects observed in the residual plots, normal plots, and $R^2$ of (4). Finally, we sought to define a metric useful in the sense of the best possible efficiency, i.e., it succinctly captured not just the performance of an individual compressor, but that it did so with respect to $E[H_\infty]$, hence the inclusion of $\min(s_{\text{comp}})$.

# 6. RESULTS
## 6.1 General Discussion
Of the 616 possible combinations (14 compressors x 44 test files), 595 test run combinations were successful (cf. Appendix A) and thus usable for results analysis. The 244 maximum compression test run results replaced the respective default compression results when we analyzed maximum compression settings. Since XGrind was run atop the CoLinux emulator (cf. Appendix B), we needed to account for timing differences. XGrind execution times, $t_{exec}$, were scaled based on GZIP taking ~1.047 times longer to run in CoLinux, versus natively in Microsoft Windows, on all tests.

## 6.2 Test System Environment
All tests were executed on a machine having the specifications given in Table 6. A 15-second pause was used between tests to provide recovery time for subsystems, e.g., virtual memory. We also randomly re-ordered all test combinations for similar reasons.

**Table 6. Compression test system specifications**

| Category | Value |
|---|---|
| Model | Gateway E Series SL2 E 6300 |
| CPU | Pentium 4 HT CPU (520) @ 3.20 GHz |
| RAM | 2 x 512 MB DDR2 DIMMs @ 400 MHz |
| Hard Disk | 1 x Western Digital Caviar, 800BD-22JMA0 |
| Hard Disk Partitions | C: 56314 MB, ~10% free, 1024 MB VM (fixed) |
| | D: 20003 MB, ~75% free, 3072 MB VM (fixed) |
| OS | Windows XP Pro (SP2), CoLinux (XGrind only) |

## 6.3 Compression Ratio and Execution Speed
The mean scores across the test corpus for each compressor with respect to (2) and (3) is shown in Figure 4. The standard deviation is denoted by the paired horizontal bars about each data point. Since (2) and (3) use base-2 and base-10 logarithms, respectively, a difference of one reflects a compressor achieves two (ten) times the compression ratio (speed). For instance, PAQ yields twice the compression ratio of GZIP, but it takes ~1000x longer to execute.

However, it is difficult, to draw any more precise conclusions by comparing mean and standard deviation of individual metrics. We assess *metric* interaction of compression ratio and execution time using our proposed combined metric, $y_{eff\_prop}$, and we determine *factor* interaction using linear regression models.

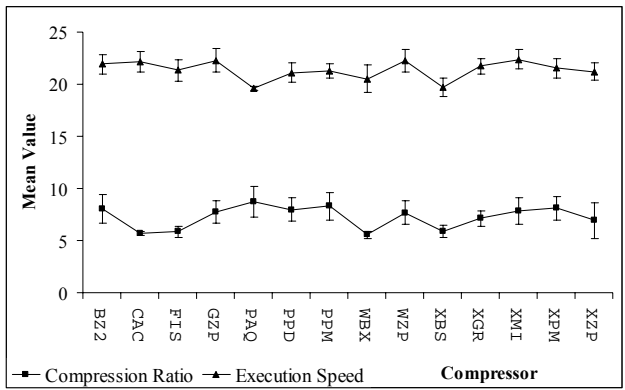

**Figure 4. Compressor vs. {compression ratio, execution speed}**

When using linear regression and analysis of variance (ANOVA), the "F-test" reflects the variance, i.e., the sum of squares of the estimated versus fitted values accounted for by the factor, where larger is better. The "Prob > F" value reflects whether the F-test is statistically significant; we used a significance level of $\alpha = 0.05$.

Table 7 shows the linear regression model used for $y_{comp\_ratio}$ (2) and $y_{comp\_speed}$ (3) using default compression. The first row shows the overall model (using all factors) is significant. The file domain (FileDom) and compressor class (ComClass) factors correspond with Tables 3 and 4, respectively. Characters (chars), lines, tree depth, $E[H_\infty]$, and $H_1$ factors are as listed in Table 2. The model validates key expectations, e.g., $E[H_\infty]$ predicts $y_{comp\_ratio}$. The run ID factor is not significant, i.e., the random run re-ordering was effective. All interaction effects are negligible and are not shown. The compressor class and $H_1$ are valid factors to predict $y_{comp\_ratio}$ and $y_{comp\_speed}$. However, the file domain is significant only with respect to $y_{comp\_ratio}$. Thus, according to this model, the selection of a compressor should be based on the compressor class and file domain, in order to maximize the compression ratio.

**Table 7. ANOVA — compression ratio and speed (default)**

| Factor | DF | $y_{comp\_ratio}$ (2) | | $y_{comp\_speed}$ (3) | |
|---|---|---|---|---|---|
| | | F-test | Prob > F | F-test | Prob > F |
| All Factors | 18 | 82.463 | < 0.0001 | 24.847 | < 0.0001 |
| Run ID | 1 | 1.184 | 0.1798 | 0.019 | 0.8895 |
| FileDom | 8 | 43.106 | <0.0001 | 2.486 | 0.0117 |
| ComClass | 4 | 568.332 | <0.0001 | 73.103 | <0.0001 |
| Chars | 1 | 18.170 | <0.0001 | 34.589 | <0.0001 |
| Lines | 1 | 0.156 | 0.6258 | 2.504 | 0.1141 |
| Depth | 1 | 0.245 | 0.5418 | 2.093 | 0.1484 |
| $H_1$ | 1 | 15.431 | <0.0001 | 21.576 | <0.0001 |
| $E[H_\infty]$ | 1 | 75.330 | <0.0001 | 2.065 | 0.1513 |
| Error | 576 | 0.656 | | 0.931 | |
| Total$_{corr}$ | 594 | | | | |

## 6.4 Combined Efficiency
To model combined efficiency metrics, we only factors that were significant with respect to predicting $y_{comp\_ratio}$ or $y_{comp\_speed}$, thus reducing the number of degrees of freedom (DF) from 18 to 15. The linear model for default compression is shown in Table 8; maximum compression results are again similar. We conclude the file domain, compressor class and $E[H_\infty]$ are the best factors to assess overall efficiency. Excepting the difference with respect to the $H_1$ factor, we *could* potentially conclude from Table 8 that the two metrics, $y_{eff\_old}$ and $y_{eff\_prop}$, are equivalent.

**Table 8. ANOVA — combined efficiency (default)**

| Factor | DF | $y_{eff\_old}$ (4) | | $y_{eff\_prop}$ (5) | |
|---|---|---|---|---|---|
| | | F-test | Prob > F | F-test | Prob > F |
| All Factors | 15 | 29.560 | <0.0001 | 34.551 | <0.0001 |
| FileDom | 8 | 3.289 | 0.0011 | 11.299 | <0.0001 |
| ComClass | 4 | 72.050 | <0.0001 | 93.324 | <0.0001 |
| Chars | 1 | 0.861 | 0.3537 | 6.382 | 0.0118 |
| $H_1$ | 1 | 14.487 | 0.0002 | 0.2240 | 0.6362 |
| $E[H_\infty]$ | 1 | 39.415 | <0.0001 | 34.058 | <0.0001 |
| Error | 579 | 4.8914 | | 0.9745 | |
| Total$_{corr}$ | 594 | | | | |

However, (5) yields a better predicted versus residual values plot, a standard method of assessing model fitness. Several visual cues in Figure 5a show (4) is fitted to a poor linear model or is a poor metric. These residuals, plotted against the vertical axis, are not equally distributed about zero, there is a noticeable diagonal slant of the residuals, and a large set of outliers are present. Figure 5b plots the residuals of our proposed efficiency metric (5); they are not significant skewed and are distributed evenly about zero with respect to the vertical axis.

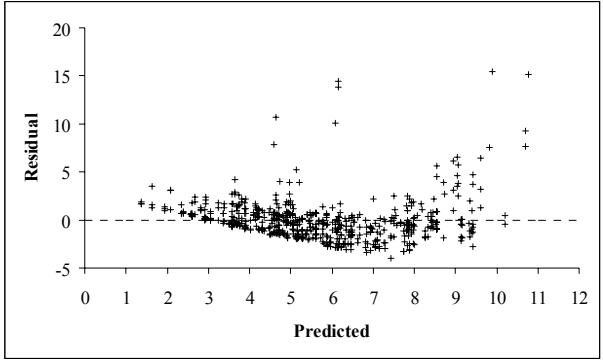

(a) Control metric — $y_{\text{eff\_old}}$ (4)

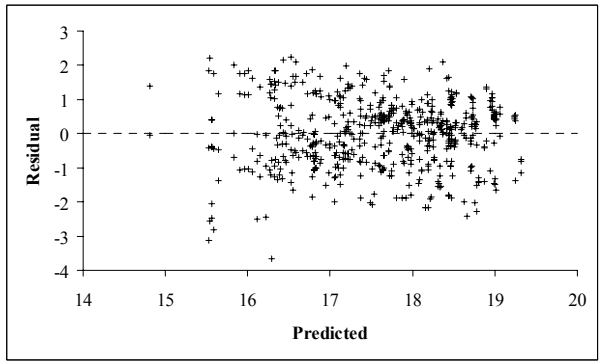

(b) Proposed metric — $y_{\text{eff\_prop}}$ (5)

**Figure 5. Predicted versus residual scores (efficiency)**

Since we have shown the proposed efficiency metric results in a suitable model, we can now attempt to assess which compressor is most efficient. The scores for each file and the compressor means are shown in Figure 6, but it is difficult using this (common) approach to determine which compressors are the most efficient.

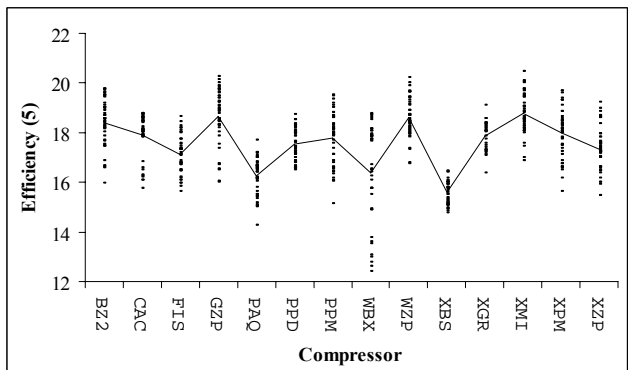

**Figure 6. Compressor versus efficiency — $y_{\text{eff\_prop}}$ (5)**

A pair-wise means comparison can determine statistically similar factors at a particular confidence level (again, $\alpha = 0.05$). After we construct the linear model for $y_{\text{eff\_prop}}$ (Table 8), we applied the Tukey-Kramer honestly significantly different (HSD) test to obtain Table 9, where columns $T_1 - T_7$ group statistically similar compressors. For instance, XMill, GZIP, WinZip®, and BZIP2 are statistically equivalent with respect to $y_{\text{eff\_prop}}$ (5). Furthermore, the slowest compressor, XBIS, and the compressor yielding the best compression ratio, PAQ, are also equivalent (and inefficient).

**Table 9. Pair-wise Tukey-Kramer HSD tests — $y_{\text{eff\_prop}}$ (5)**

| Compressor | $T_1$ | $T_2$ | $T_3$ | $T_4$ | $T_5$ | $T_6$ | $T_7$ | Mean |
|---|---|---|---|---|---|---|---|---|
| XMILL | X | | | | | | | 18.787 |
| GZIP | X | X | | | | | | 18.644 |
| WINZIP | X | X | | | | | | 18.614 |
| BZIP2 | X | X | X | | | | | 18.407 |
| XMLPPM | | X | X | X | | | | 18.003 |
| CACM3 | | | X | X | | | | 17.904 |
| XGRIND | | X | X | X | X | | | 17.880 |
| PPMZ2 | | | X | X | X | | | 17.780 |
| PPMD | | | | X | X | | | 17.543 |
| XMLZIP | | | | X | X | | | 17.350 |
| FIS | | | | | X | | | 17.130 |
| WBXML | | | | | | X | | 16.366 |
| PAQ | | | | | | X | X | 16.272 |
| XBIS | | | | | | | X | 15.576 |

From these results, we conclude XMill offers the best combined efficiency, $y_{\text{eff\_prop}}$, and is statistically similar to WinZip®, GZIP, and BZIP2. XMLZIP is hindered by its tendency to increase file size. WinZip® increased one file by two bytes; no other increases were observed. The new PPMd algorithm integrated in WinZip® is an average compressor (for this corpus and metric). The binary formats, Fast Infoset, WBXML, and XBIS, have a low efficiency.

### 6.5 File Size Categorization

We observe some performance differences if we filter the results by uncompressed file size. Notably, WBXML achieves higher efficiency on small files, i.e., those less than ~6 KB. It is also the fastest compressor of these files and achieves compression on the order of $H_1$. Given WBXML has low computational overhead and its widespread use, this result warrants further investigation. Other XML binary formatters did not share WBXML's execution speed on small files. Ironically, WBXML is the slowest compressor on larger test files; we were unable to determine if this effect is an implementation or algorithm-specific issue.

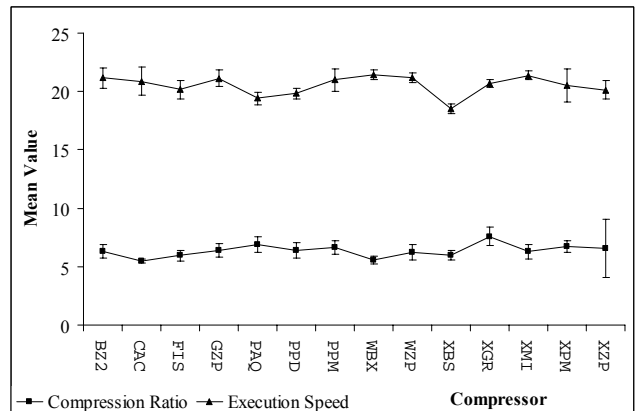

**Figure 7. Compressor vs. {compression ratio, speed} (< 6 KB)**

Figure 8 plots the compression ratio for each corpus file listed in Table 2, as yielded by the most efficient compressor, XMill, and bounded by the compression ratios yielded by the CACM3 and PAQ compressors, the compressors used to obtain $H_1$ and $E[H]_\infty$, respectively. The logarithmic horizontal axis visually reinforces why we used a logarithmic transform in the compression metrics; we otherwise would have been unable to apply linear regression.

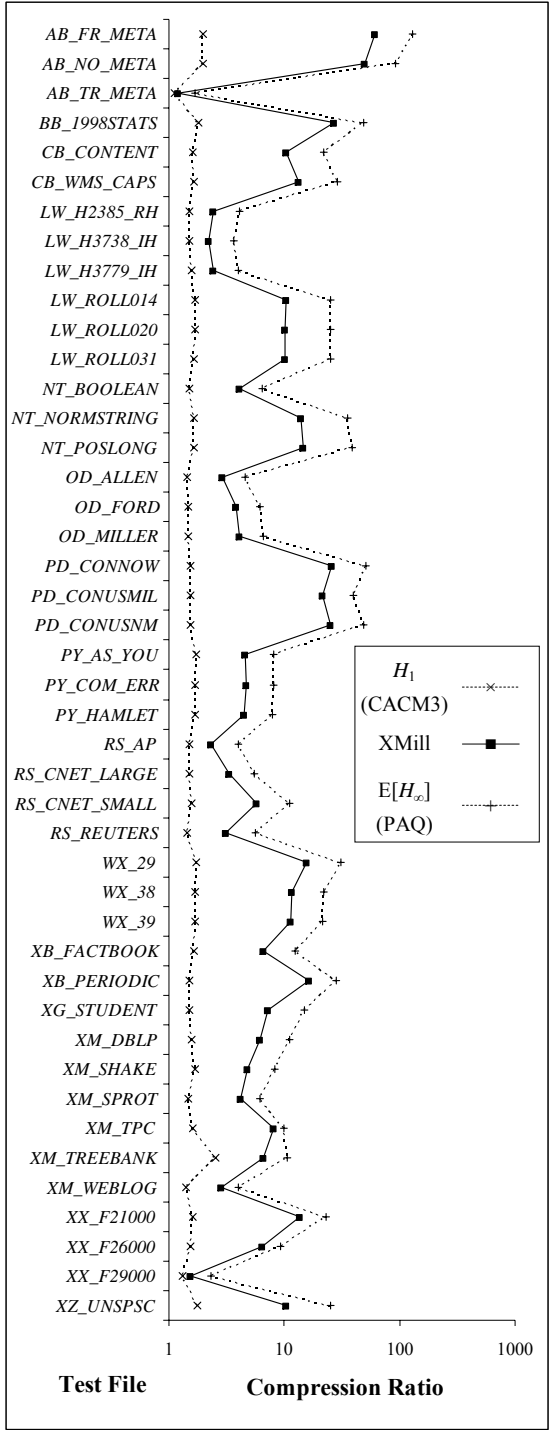

**Figure 8. Compression ratio of XMill (most efficient compressor with respect to $y_{eff\_prop}$)**

# 7. CONCLUSION

We proposed a corpus of XML test files for assessing compressor performance and a combined efficiency metric, $y_{eff\_prop}$, to assess compression ratio and speed simultaneously and then used linear regression to rank the compressors with respect to the proposed metric. For readers interested in applying the results of this study, we recommend the following courses of action:

1. In most instances, a general-purpose compressor, e.g., a zip utility, should be used. If maximum parsing and compression speed is needed in an XML-intensive application, compressors such as XMill may be useful. This is contingent on the benefit gained over using native XML along with general-purpose compressors.

2. The tested binary formatters often compress small XML files well. Given the plethora of existing binary formats, e.g., WBXML, FIS, and MPEG-7, along with the EXIWG's efforts, other binary formats may not be needed. Our results indicate that binary formats, e.g., WBXML, are best applied to small files. This should be considered and verified as the EXIWG prepares their binary format specification.

Several avenues of research were not investigated in this study; we suggest future work explore one or more of the following:

1. The EXIWG recently identified Efficient XML as the basis of its binary format [15]; it and other recently developed binary formats or compressors should be tested in future work. Although a publicly accessible version of Efficient XML was not available during this study (cf. Section 4.4), a development kit has since been released on the vendor's website [16].

2. Neither decompression performance nor memory requirements were assessed in this study. In addition to randomizing file and compressor combinations, the use of two disks would also reduce system issues, e.g., fragmentation. We also conducted only one run of the test file and compressor combinations—repeated trials would enable determining confidence intervals.

3. In addition to discussing the validity and utility of the proposed XML corpus and combined efficiency metric, it would be useful to explore whether 2-stage compression is useful, e.g., applying a pre-processing transform such as XMill. This may lead to integrating an XML model in a general-purpose compressor and aid in limiting proliferation of proprietary solutions.

In sum, we have motivated the utility of an XML test file corpus, akin to the Canterbury and Calgary corpora, along with the utility of a combined efficiency metric for assessing compression ratio and compression speed simultaneously. We have also provided sufficient detail to enable experimental repeatability and relative to studies that only compare means, we have shown the utility of using linear regression for analyzing compressor performance.

## Acknowledgement
We thank Andy Andrews for his suggestions and encouragement while conducting this research. We also thank him for keeping us abreast of recent developments in the XML domain.

## APPENDIX A: Uncollected Test Samples

The compressor and test file combinations listed in Table 10 are those that failed to complete. For the most part, these test failures were attributed to parser errors in the compressor being tested.

However, WBXML failures were due to excessive execution time; even after being granted several *days* of execution, WBXML had not compressed the files listed in Table 10. Given that WBXML is used with wireless devices, this may be acceptable; however, this issue warrants further investigation.

**Table 10. Failed compressor / corpus test file combinations**

| Compressor | Test File |
|---|---|
| WBXML | ab_FR_meta<br>ab_NO_meta<br>xb_factbook<br>xz_UNSPSC-2 |
| XBIS | ab_FR_meta<br>cb_content<br>xm_treebank |
| XGRIND | ab_FR_meta[1]<br>ab_NO_meta[2]<br>cb_content<br>cb_wms_caps<br>lw_h* (all)<br>lw_roll* (all)[2]<br>nt* (all)<br>wx* (all)<br>xb_periodic<br>xm_dblp<br>xm_sprot |

## APPENDIX B: CoLinux Experiences

This appendix briefly summarizes how we installed the CoLinux emulator [10] of the Linux operating system (cf. Sections 4.3.1 and 6.1), whose capabilities are similar to cygwin and VMWare, two other ways of accessing Linux from a Windows environment. Although CoLinux runs as an application in Microsoft Windows®, its installation can be quite challenging [10, 11, 12, 39].

First, CoLinux requires an "installed" Linux distribution; we used a 1+ GB compressed Debian variant, of the many available at the CoLinux site [10]. Installing CoLinux on a Windows XP® system may require disabling of some memory protection features and system reboot after applying the "/NoExecute=AlwaysOff" setting to the "boot.ini" file. This modification is typically used if a hard system crash occurs when CoLinux is launched.

Since Windows XP® retains control of any physical devices, e.g., disk drives, a convenient method to share data between CoLinux and Windows XP® is via a network link. This is readily achieved by using a network bridge in Windows XP®; a bridge should be disabled when CoLinux is not in use, especially prior to a system restart, as it may prevent subsequent logons.

We used the file-sharing service Samba, which required a registry modification to Windows XP® and a download via "apt-get" in CoLinux (once the network bridge is established). Alternatively, an FTP server could be installed on the Windows XP® system.

---

[1] Indicates failure during max (not default) compression testing.

## APPENDIX C: Corpus Test File Sources

We collected corpus files from several sources (cf. Section 5.1), with a broad goal of spanning, as much as possible, the domains and sizes of XML files in common use. As with any benchmark, these files are only one set of test files that could be used to assess a system's performance. A key source was sample files packaged with XML compressors. We then added XML-formatted versions of files used in other compression studies or in other corpora. We also added files of interest to our research, along with additional ones to span the file sizes we determined to be relevant.

Since copyright restrictions may preclude us from providing the native or modified XML files, we provide the source locations of all test files used in this study in Table 11. Since pre-processing steps used by researchers may vary, the results may differ slightly between experiments; in addition to providing source locations, Table 11 also illustrates the scope of files in the corpus. The first two letters of each file's name serves as its prefix (cf. Table 2).

**Table 11. Corpus test file source locations**

| Prefix | File Source Location (URL) |
|---|---|
| AB | http://air-climate.eionet.eu.int/databases/airbase/airbasexml/index_html#downld |
| BB | http://www.ibiblio.org/xml/examples/ |
| CB | http://www.cubewerx.com/main/cwxml/ |
| LW | http://xml.house.gov/ |
| NT | http://xw2k.sdct.itl.nist.gov/brady/xml/ |
| OD | http://www.oracle.com/technology/tech/xml/xmldb/index.html |
| PD | http://www.kensall.com/gov/perdiem/ |
| PY | 1. http://www.ibiblio.org/xml/examples/shakespeare/<br>2. http://www.oasis-open.org/cover/bosakShakespeare200.html |
| RS | 1. http://hosted.ap.org/dynamic/fronts/RSS_FEEDS?SITE=AP<br>2. http://www.cnet.com/4520-6022-5115113.html<br>3. http://reviews.cnet.com/4924-5_7-0.xml?7eChoice=1&orderBy=-7rvDte&maxhits=50000000<br>4. http://today.reuters.com/rss/newsrss.aspx |
| WX | 1. http://weather.gov/xml/<br>2. http://weather.gov/forecasts/xml/SOAP_server/ndfdXML.htm<br>3. http://weather.gov/forecasts/xml/SOAP_server/ndfdSOAPByDay.htm |
| XB | http://xbis.sourceforge.net/ |
| XG | http://sourceforge.net/projects/xgrind/ |
| XM | http://sourceforge.net/projects/xmill |
| XX | http://www.ictcompress.com/downloadxml.html |
| XZ | http://www.xmls.com —briefly, also at http://aslam.szabist.edu.pk/XML+Solution/product/ xml_zip.html |